\documentclass[preprint,12pt]{cas-sc}
\usepackage[authoryear]{natbib}
\usepackage{lineno}

\def\tsc#1{\csdef{#1}{\textsc{\lowercase{#1}}\xspace}}
\tsc{WGM}
\tsc{QE}

\begin{document}
\let\WriteBookmarks\relax
\def\floatpagepagefraction{1}
\def\textpagefraction{.001}

\shorttitle{Venus thermal wave structure}    

\shortauthors{Giles, Greathouse, Irwin, Encrenaz and Brecht}  

\title [mode = title]{Three-dimensional structure of thermal waves in Venus' mesosphere from ground-based observations}  

\author[1]{Rohini S. Giles}[type=editor,
                        auid=000,bioid=1,
                        orcid=0000-0002-7665-6562]
\cormark[1]
\ead{rohini.giles@swri.org}

\author[1]{Thomas K. Greathouse}
\author[2]{Patrick G. J. Irwin}
\author[3]{Thérèse Encrenaz}
\author[4]{Amanda Brecht}

\affiliation[1]{organization={Space Science and Engineering Division, Southwest Research Institute},
            city={San Antonio},
            state={Texas},
            country={USA}}
\affiliation[2]{organization={Department of Physics, University of Oxford},
            city={Oxford},
            country={UK}}
\affiliation[3]{organization={LEISA, Observatoire de Paris},
            city={Paris},
            country={France}}
\affiliation[4]{organization={NASA Ames Research Center},
            city={Moffett Field},
            state={California},
            country={USA}}

\cortext[1]{Corresponding author}


\begin{abstract}
High spectral resolution observations of Venus were obtained with the TEXES instrument at NASA's Infrared Telescope Facility. These observations focus on a CO\textsubscript{2} absorption feature at 791.4 cm\textsuperscript{-1} as the shape of this absorption feature can be used to retrieve the vertical temperature profile in Venus' mesosphere. By scan-mapping the planet, we are able to build up three-dimensional temperature maps of Venus' atmosphere, covering one Earth-facing hemisphere and an altitude range of 60--83 km. A temperature map from February 12, 2019 clearly shows the three-dimensional structure of a planetary-scale thermal wave. This wave pattern appears strongest in the mid-latitudes of Venus, has a zonal wavenumber of 2--4 and the wave fronts tilt eastward with altitude at an angle of 8--15 degrees per km. This is consistent with a thermal tide propagating upwards from Venus' upper cloud decks. Ground-based observations provide the opportunity to study Venus' temperature structure on an ongoing basis. 
\end{abstract}


\begin{highlights}
\item Ground-based observations were used to map Venus’ three-dimensional thermal structure
\item A temperature offset map from February 2019 shows a planetary-scale wave
\item The wave pattern is consistent with an upward propagating thermal tide
\end{highlights}


\maketitle



\section{Introduction}
\label{sec:introduction}

Venus' upper cloud layer absorbs more than half of the incident solar radiation~\citep{gierasch97} and this localized energy deposit drives large-scale gravity waves known as thermal tides~\citep{sanchez-lavega17}. These tides are thought to be sun-synchronous, with an eastward phase velocity, and propagate both upwards and downwards from the forcing region in the upper cloud deck where the maximum amount of solar energy is absorbed. This movement of eastward momentum vertically away from the cloud deck region leads to the westward acceleration of atmosphere at the cloud level, which may be an important element in maintaining Venus' extreme cloud-top superrotation~\citep{fels74,newman92,lebonnois16,kouyama19}. 

Thermal tides were first observed by \cite{apt80} using ground-based mid-infrared images obtained in 1977--1979. These observations showed a strong solar-fixed component to the thermal emission, with diurnal (zonal wavenumber of 1), semidiurnal (zonal wavenumber of 2) and higher order components. Further observations of the thermal tides have been obtained from instruments on missions to Venus, including the Orbiter Infrared Radiometer (OIR) on Pioneer Venus~\citep{taylor80,schofield83b} and the Fourier Transform Spectrometer (FTS) on Venera 15~\citep{zasova07}, both of which showed clear diurnal and semidiurnal components. \cite{zasova07} found that semidiurnal component dominates in the region immediately above the cloud top (58--90 km) at low latitudes, while the diurnal component dominates at higher altitudes at low latitudes and at the cloud-top level in the polar regions. The most recent observations of the thermal tides are from the Akatsuki spacecraft, which is currently in orbit around Venus; global observations from the Longwave Infrared Camera (LIR) show a clear tidal structure that remains fixed in local time over a two year period~\citep{kouyama19}. 

Two previous studies using spacecraft observations have shown that the phase of the thermal tides changes with altitude, which is indicative of vertical propagation. The OIR instrument on Pioneer Venus consisted of 10 different channels, each of which probed a different altitude in Venus' atmosphere. \cite{schofield83b} compared the semidiurnal wave pattern observed in five of these channels and found that the phase shifted eastward with altitude. The LIR instrument on the Akatsuki spacecraft consists of a single wavelength filter, but \cite{kouyama19} and \cite{akiba21} used the fact that observations obtained at different emission angles probe slightly different altitudes in order to similarly show that there is an eastward shift with altitude.

We seek to build on these previous studies by using high spectral resolution ground-based observations of Venus to produce vertically-resolved three-dimensional temperature maps of the planet's middle atmosphere. These maps can be used to study a range of thermal phenomena, including the thermal tide structure. In this paper, we present initial observations from February 12, 2019, which show a clear vertically-propagating planetary-scale wave. The observations and radiative transfer modeling are described in Section~\ref{sec:observations}, and the wave structure is described in Section~\ref{sec:results} and discussed in Section~\ref{sec:discussion}. This initial observation forms part of an ongoing long-term monitoring campaign to study how Venus' temperature structure varies with time and future studies will combine observations obtained over several years. 

\section{Observations}
\label{sec:observations}

\subsection{TEXES observations}

\begin{figure}
\centering
\includegraphics[width=7cm]{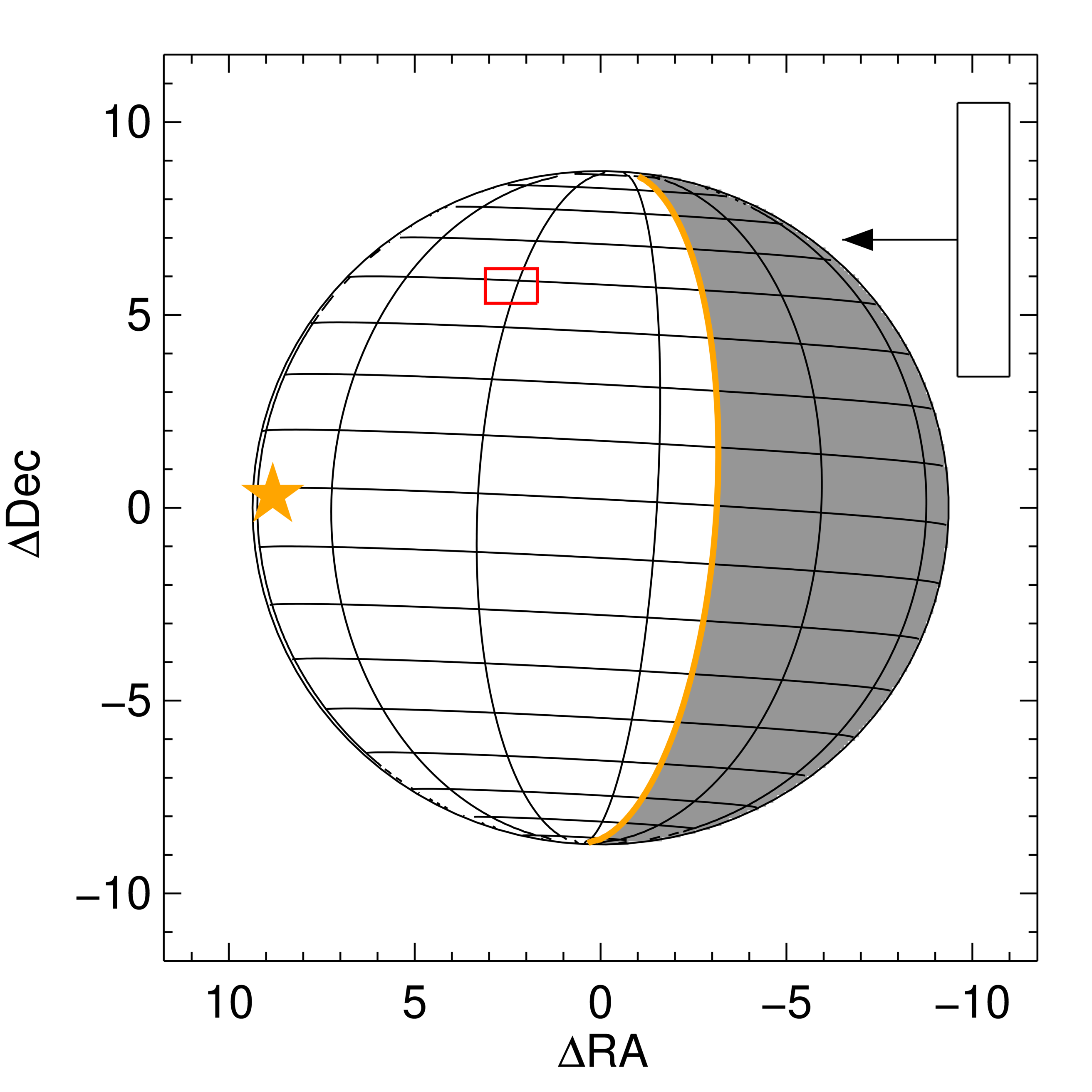}
\caption{The geometry of the Venus observations on February 12, 2019. The orange star marks the location of the sub-solar point and the orange line marks the dawn terminator. The regions of the planet shown in white are on the dayside and the regions of the planet shown in grey are on the nightside. The TEXES slit dimensions and step direction are shown in the upper right corner. The spatial resolution in ideal observing conditions is shown by the red rectangle.}
\label{fig:geometry}
\end{figure}

In February 2019, high spectral resolution observations of Venus were carried out using the Texas Echelon Cross Echelle Spectrograph (TEXES), mounted at NASA's 3-m Infrared Telescope Facility (IRTF) in Hawaii. Observations were made on six days within a single observing run, but in this paper we focus exclusively on February 12, which had the best observing conditions. TEXES~\citep{lacy02} is a long-slit cross-dispersed grating spectrograph that covers wavelengths of 4.5--25 \textmu m in the mid-infrared, with a resolving power between 4,000 and 100,000 depending on the instrument mode and wavelength. On February 12, 2019, Venus had an apparent diameter of 17.5'' and the dawn side of the planet was visible. The observing geometry is shown in Figure~\ref{fig:geometry}, which shows Venus as it appeared on the sky. The orange star marks the location of the sub-solar point and the orange line marks the dawn terminator.

\begin{figure}
\centering
\includegraphics[width=14cm]{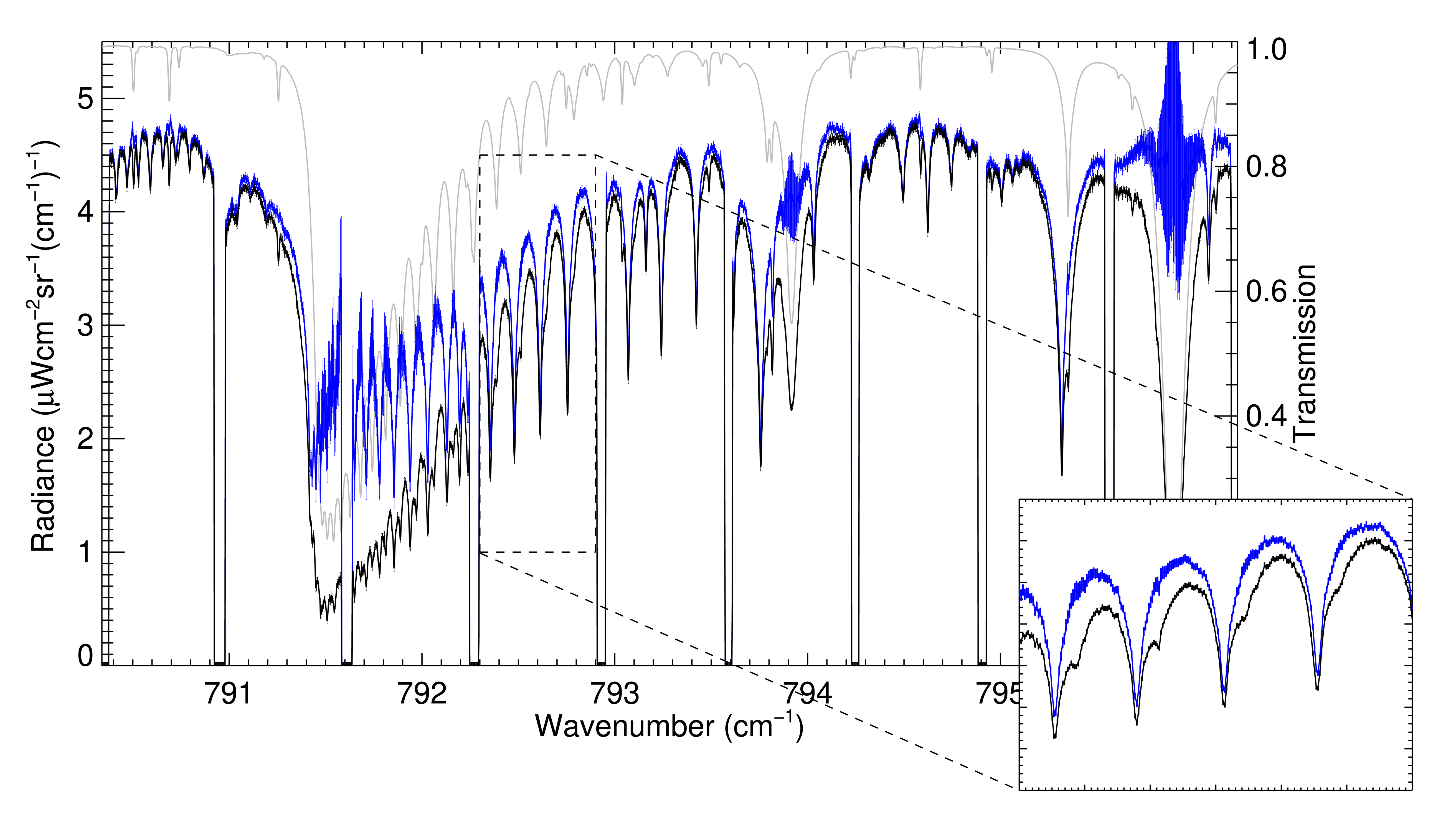}
\caption{An example of the TEXES observations of Venus in the 793 cm\textsuperscript{-1} CO\textsubscript{2} spectral setting, obtained on February 12, 2019. The TEXES spectrum prior to telluric correction is shown in black. The regularly spaced small gaps in the spectrum (where the radiance equals zero) are due to the small coverage gaps between spectral orders. The grey line shows the best-fit telluric transmission spectrum obtained from fitting the telluric lines. The blue line and associated error bars show the spectrum after it has been divided by the best-fit telluric spectrum.}
\label{fig:spectrum_correction}
\end{figure}

In order to study Venus' mesospheric temperatures, we used TEXES to observe the strong 11101$\leftarrow$10002 CO\textsubscript{2} Q-branch absorption feature at 791.4 cm\textsuperscript{-1} (12.6 \textmu m). The observations were centered at $\sim$793 cm\textsuperscript{-1}, with a spectral bandpass of $\sim$5.5 cm\textsuperscript{-1}. At the center of this CO\textsubscript{2} band, telluric transmission is $\sim$0.5 but the Venus absorption feature is both broader and deeper and can therefore still be observed. TEXES was used in its highest spectral resolution mode (R=72,000 at this spectral setting), which allows both the telluric and Venusian spectral lines to be resolved. Because the CO\textsubscript{2} abundance in Venus' atmosphere is well-known and homogeneous, the shape of the absorption feature is governed by the vertical temperature profile of the atmosphere. At the center of the band, the gaseous opacity is highest and so higher altitudes are probed, whereas the edges of the band and the continuum, where the absorption is weaker, probe deeper into the atmosphere. Together, this spectral interval probes the 60--83 km altitude region. An example of the spectrum obtained by TEXES is shown in Figure~\ref{fig:spectrum_correction}. The majority of the strong absorption lines in the spectrum are from CO\textsubscript{2} and the TEXES data prior to telluric correction (black) shows both telluric and Venusian CO\textsubscript{2} absorption with a small offset between the two due to the Doppler shift. 

Three-dimensional spectral image cubes of Venus were obtained by using the TEXES scan-map observing mode. The 1.4''-wide TEXES entrance slit was aligned along the celestial north-south direction and it was scanned across the planet from west to east, in 0.7'' increments. The size of the entrance slit and the scan direction are shown in the upper right corner of Figure~\ref{fig:geometry}. Allowing for several steps of clear sky at the beginning and end of each scan, a total of 38 steps were required to cover the planet in the east-west direction. In the scan-map observing mode, each step has a defined 2 s integration time. To achieve a total integration time of 8 s for each position on the planet, we repeated each scan 4 times. At 793 cm\textsuperscript{-1}, we use an entrance slit that is 7.1'' long; this is the longest slit that can be used without adjacent echelon orders overlapping on the detector. On February 12, 2019, Venus had an apparent diameter of 17.5'', so a total of 4 different north-south offset positions were required to obtain full coverage with some overlap between adjacent positions. The total integration time required for the observing sequence was therefore $38\times2\times4\times4$ = 1216 s = 20.3 min.

The red square in Figure~\ref{fig:geometry} shows the spatial resolution in ideal observing conditions. The slit width of 1.4'' is the maximum spatial resolution that can be obtained in the across-slit (east-west) direction and the step size of 0.7'' was chosen to ensure Nyquist sampling. The maximum spatial resolution in the along-slit (north-south) direction is defined by the diffraction limit at the IRTF, which is 0.9'' at 793 cm\textsuperscript{-1}. The plate scale along the slit is 0.36'', so the sampling interval is slightly better than Nyquist sampling. In reality, the spatial resolution in both directions is also affected by the atmospheric seeing, which can be high during the daytime hours when Venus is observable, and telescope drift. In this paper, we focus on February 12, 2019 because the observing conditions were good and the spatial resolution was therefore close to ideal. 

\subsection{Data reduction}
\label{sec:reduction}

The individual TEXES Venus scans were first reduced using the data reduction pipeline described in \cite{lacy02}. The data were flat-fielded, sky-subtracted and corrected for instrumental geometric optical distortions. The observations were wavenumber-calibrated using telluric absorption lines and were radiometrically calibrated by using the measured radiance of a room temperature blackbody that was automatically placed in front of instrument aperture prior to each set of observations.

Typically, the data reduction pipeline also uses the sky observations to correct for the effect of telluric absorption. However, for the spectral setting used in this work, the telluric absorption from CO\textsubscript{2} is sufficiently high that we developed an alternative method of correcting for the Earth's atmosphere, adapted from the method described by~\cite{rudolf16} for reducing VLT/X-Shooter spectra. We used the line-by-line radiative transfer code LBLRTM~\citep{clough05} to model the transmission through the Earth's atmosphere. The reference atmosphere inputs for this code were obtained from the Global Data Assimilation System~\cite[GDAS,][]{rodell04} which produces global maps of the Earth's temperatures and H\textsubscript{2}O relative humidity between sea level and an altitude of 26 km every 3 hours; we selected the location of the observatory and the times closest to the observations. Other atmospheric molecular abundances, along with temperatures and H\textsubscript{2}O abundances above 26 km, were obtained from the MIPAS equatorial model atmosphere~\citep{mipas_model} and the CO\textsubscript{2} profile was scaled to match the mean monthly CO\textsubscript{2} abundance measured at Mauna Loa~\citep{mauna_loa_co2}. This atmospheric data was used to generate a telluric transmission spectrum using the LBLRTM code. We assume that the CO\textsubscript{2} abundance is well-defined, but as H\textsubscript{2}O varies rapidly, we iteratively adjust the H\textsubscript{2}O abundance to best fit the strong H\textsubscript{2}O absorption line at 796 cm\textsuperscript{-1}, which is located in a spectral region with minimal CO\textsubscript{2} absorption. The Venus observations were then divided by this best-fit telluric transmission in order to obtain a `clean' spectrum. This process can be seen in Figure~\ref{fig:spectrum_correction}. The spectrum prior to telluric correction is shown in black, and the corrected spectrum is shown in blue. The insert shows a zoomed-in portion of the spectrum. The black line shows the larger Venusian CO\textsubscript{2} absorption lines, which have been Doppler-shifted relative to the adjacent smaller telluric CO\textsubscript{2} absorption lines. The combination of the two gives an asymmetric appearance to each absorption line. In contrast, the corrected line (blue) shows a smooth curve between each Venusian line. 

The wavenumber- and radiometrically-calibrated data were then geometrically-calibrated. The plate scale, slit orientation, step size and step direction were used to calculate the shape of the planet's limb. This calculated limb was then manually adjusted to fit the actual observed limb for each scan. This allowed each pixel to be assigned a latitude and longitude on Venus, and the latter was converted into a local solar time. The spectrum from each pixel was Doppler shifted into the rest frame and then projected onto a latitude-local time grid with a bin size of 5$^{\circ}$ in the latitude direction and 0.5 hrs in the local time direction. The multiple north-south offset scans were co-added to produce complete latitudinal coverage. The final re-projected and co-added data cube from February 12, 2019 has continuous coverage of latitudes between 60$^{\circ}$S and 60$^{\circ}$N and local times between 3:30 AM and 11:00 AM. The number of 0.36'' $\times$ 1.4'' (plate scale $\times$ slit width) pixels that are co-added into each bin varies from 1 at the highest latitudes, to 25 at the low latitudes that comprise the overlap between adjacent north-south offsets. 

The error bars on the final co-added spectra were calculated by adding in quadrature the error on the TEXES spectrum prior to telluric correction and the error on the best-fit telluric transmission spectrum. The error on the TEXES spectrum is calculated by taking the difference between the radiances observed in adjacent wavenumber bins and calculating the local ($\pm$0.1 cm\textsuperscript{-1}) standard deviation of these differences; we find the region of the spectrum where the local standard deviation is a minimum (where the spectrum most closely approximates a straight line) and use this value as the absolute error. The black line in Figure~\ref{fig:spectrum_correction} shows the pre-telluric correction spectrum and associated small error bars from a single pixel; when multiple spectra are co-added the error bars would be even smaller. The telluric transmission error is calculated by assuming a 10\% error on both the CO\textsubscript{2} and H\textsubscript{2}O abundances, running the telluric model with these altered values and then adding the two errors in quadrature. The blue line in Figure~\ref{fig:spectrum_correction} shows the combined error bars on the spectrum after telluric correction. In the continuum regions, the small error from the pre-telluric correction spectrum dominates but in regions of telluric absorption (e.g. the H\textsubscript{2}O line at 795.9 cm\textsuperscript{-1}), the telluric error dominates.

\subsection{Spectral modeling}
\label{sec:spectral_modeling}

In order to retrieve the vertical temperature profiles, the IRTF/TEXES spectra were modeled using NEMESIS \cite[Non-Linear Optimal Estimator for Multivariate Spectral Analysis,][]{irwin08}. NEMESIS is a radiative transfer and retrieval tool that has been applied to a wide range of planetary atmospheres and has previously been used to model Venus' atmosphere using near-infrared observations from the Visible and Infrared Thermal Imaging Spectrometer (VIRTIS) on Venus Express~\citep{tsang08b} and mid- to far-infrared observations from Pioneer Venus OIR and Venera 15 FTS~\citep{koukouli05}. NEMESIS consists of a radiative transfer code that computes the emergent radiation for a given atmospheric profile and an optimal estimation retrieval algorithm which iteratively determines the best-fit atmospheric parameters for an observed spectrum.

For this study, Venus' atmosphere was divided into 200 levels between the surface (92 bar) and an altitude of 150 km ($7\times10^{-11}$ bar), equally spaced in log(p). The vertical distribution of atmospheric gases and the a priori vertical temperature profile were obtained from NASA's Planetary Science Generator~\citep{villanueva18}, which in turn bases its profile primarily on the Venus International Reference Atmosphere~\cite[VIRA,][]{seiff85}, with updates from \cite{ehrenreich12},~\cite{vandaele20} and~\cite{bierson20}. The gases which absorb within the 790--796 cm\textsuperscript{-1} spectral range are CO\textsubscript{2}, SO\textsubscript{2}, H\textsubscript{2}O, HCl, O\textsubscript{3}, NO, ClO, OCS and H\textsubscript{2}S. The line data for these gases were obtained from the HITRAN molecular database~\citep{gordon22}. For CO\textsubscript{2}, we use the self-broadened line widths and we use a sub-Lorentzian lineshape~\citep{perrin89}. For all other gases, we assume a Voigt lineshape. SO\textsubscript{2}, OCS and HCl have CO\textsubscript{2}-broadened widths available in HITRAN~\citep{wilzewski16} and so these values are used; for the remaining gases, we use the air-broadened widths. Collision-induced absorption for CO\textsubscript{2} is minimal in the 790--796 cm\textsuperscript{-1} spectral range~\citep{wordsworth10}. NEMESIS makes use of the correlated-k approximation~\citep{lacis91} in order to reduce the computation time required. The k-distributions produced for each gas were convolved with a Gaussian of width 0.011 cm\textsuperscript{-1} in order to match the TEXES spectral resolution and instrument function.

Although scattering effects are significant for observations of Venus in the near-infrared windows~\citep{tsang08b}, \cite{koukouli05} showed that these effects are minimal at wavenumbers less than 900 cm\textsuperscript{-1} and it is therefore sufficient to treat Venus' clouds as pure absorbers in the 790--796 cm\textsuperscript{-1} spectral range used in this paper. As we are using a very narrow spectral interval, we also assume that the absorption cross-section is constant within this interval. We use the same Venusian cloud structure as \cite{lee12}, which consists of a cloud with constant extinction below 60 km, above which the extinction drops off with a constant scale height. We use a fixed scale height of 3.8 km, which was the best-fit value at low latitudes found by \cite{lee12}.

Using the a priori reference atmosphere, the NEMESIS radiative transfer code calculates a synthetic top-of-atmosphere spectrum. The retrieval segment of the code then compares this synthetic spectrum to the observed TEXES spectrum and the atmospheric profiles are iteratively adjusted to fit the observed spectrum, following an optimal estimation approach~\citep{rodgers00}. The code seeks to minimize a cost function which combines departure from the measurement and departure from the a priori atmospheric profile, weighted by their respective errors. In this case, we allow two elements of the atmospheric profile to vary in the retrievals: the vertical temperature profile and the cloud opacity. The cloud opacity is allowed to vary via a single scaling factor, which controls the overall opacity and is therefore equivalent to the cloud top altitude parameter in \cite{lee12}. In contrast, the temperature profile, which is the focus of this study, is allowed to vary continuously.

The measurement error consists of both the error on the corrected spectrum described in Section~\ref{sec:reduction} and a modeling error due to inaccuracies in the model (e.g., line data), which we conservatively assign as 5\% of the continuum radiance. We assume that the a priori temperature profile has an error of 25 K, which is the maximum variability observed by~\cite{rengel08}. The temperatures of each of the 200 atmospheric layers are allowed to vary independently to achieve the best fit, with a correlation length of 1.5 scale heights ($\sim$7.5 km) used to prevent unphysically sharp deviations between adjacent levels. The a priori error on the cloud opacity scaling factor is set to a large value (100\%) in order to allow this secondary parameter to vary freely.

\begin{figure}
\centering
\includegraphics[width=14cm]{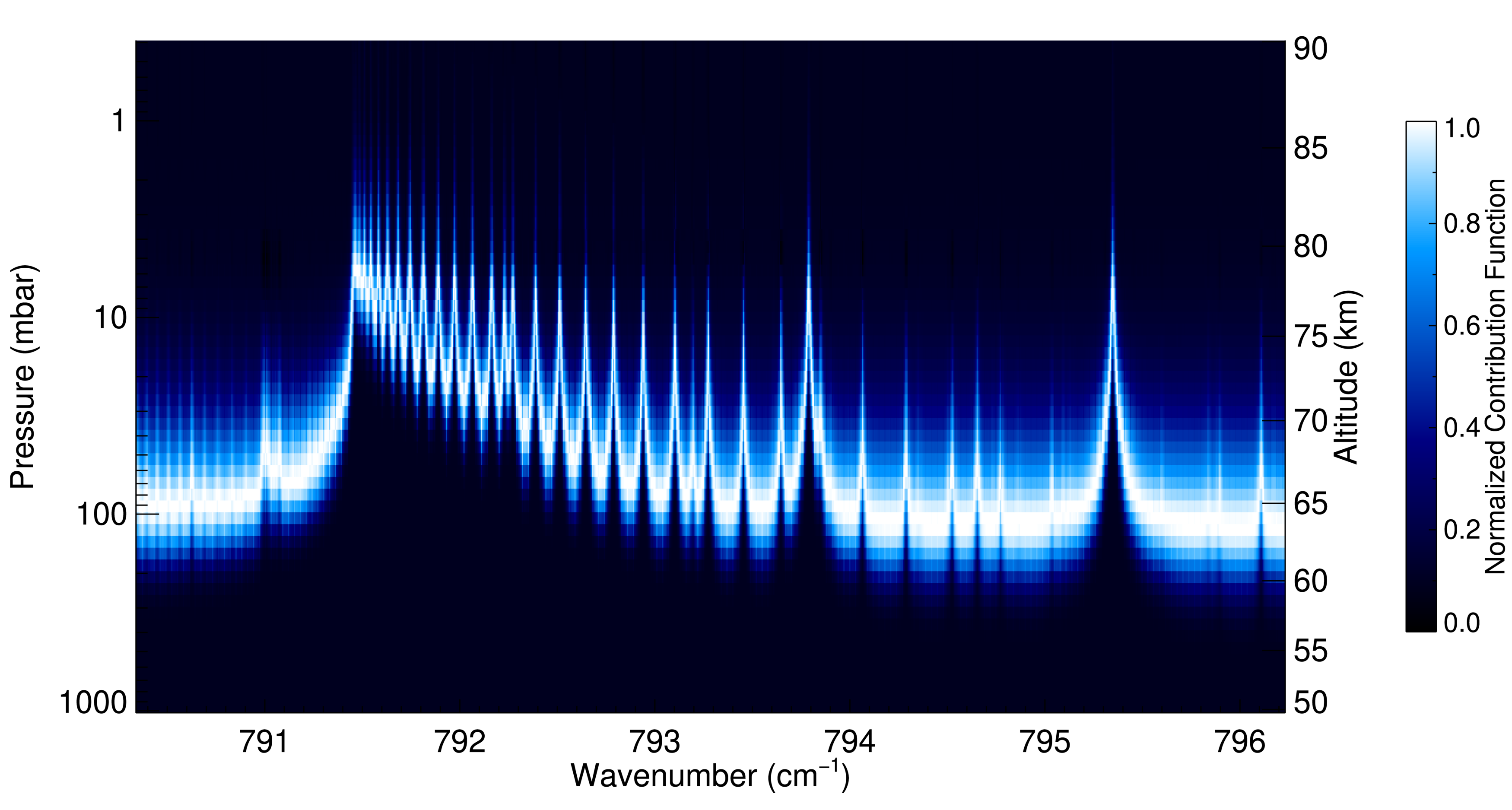}
\caption{The normalized contribution functions (temperature functional derivatives) for the spectral interval used in this study, showing where the sensitivity peaks as a function of wavenumber. In the center of the CO\textsubscript{2} absorption band at 791.4 cm\textsuperscript{-1}, the sensitivity peaks at $\sim$79 km, while in the continuum regions, the sensitivity peaks deeper in the atmosphere at $\sim$63 km.}
\label{fig:kk}
\end{figure}

Figure~\ref{fig:kk} shows the normalized contribution functions (temperature functional derivatives) for the spectral interval used in this study. The contribution functions show the extent to which a change in the temperature of a given atmospheric layer affects the observed top-of-atmosphere radiance; for each wavenumber, the altitude level where the contribution function is the highest is the pressure of maximum sensitivity. In the center of the CO\textsubscript{2} absorption band at 791.4 cm\textsuperscript{-1}, the sensitivity peaks at $\sim$79 km, while in the continuum regions, the sensitivity peaks deeper in the atmosphere at $\sim$63 km. Overall, the combined 790--796 cm\textsuperscript{-1} spectral region is sensitive to the 60--83 km altitude range, which is within and immediately above the Venus' cloud layer. Based on the width of the contribution functions, the vertical resolution in this altitude range is 4--9 km.

\begin{figure}
\centering
\includegraphics[width=12cm]{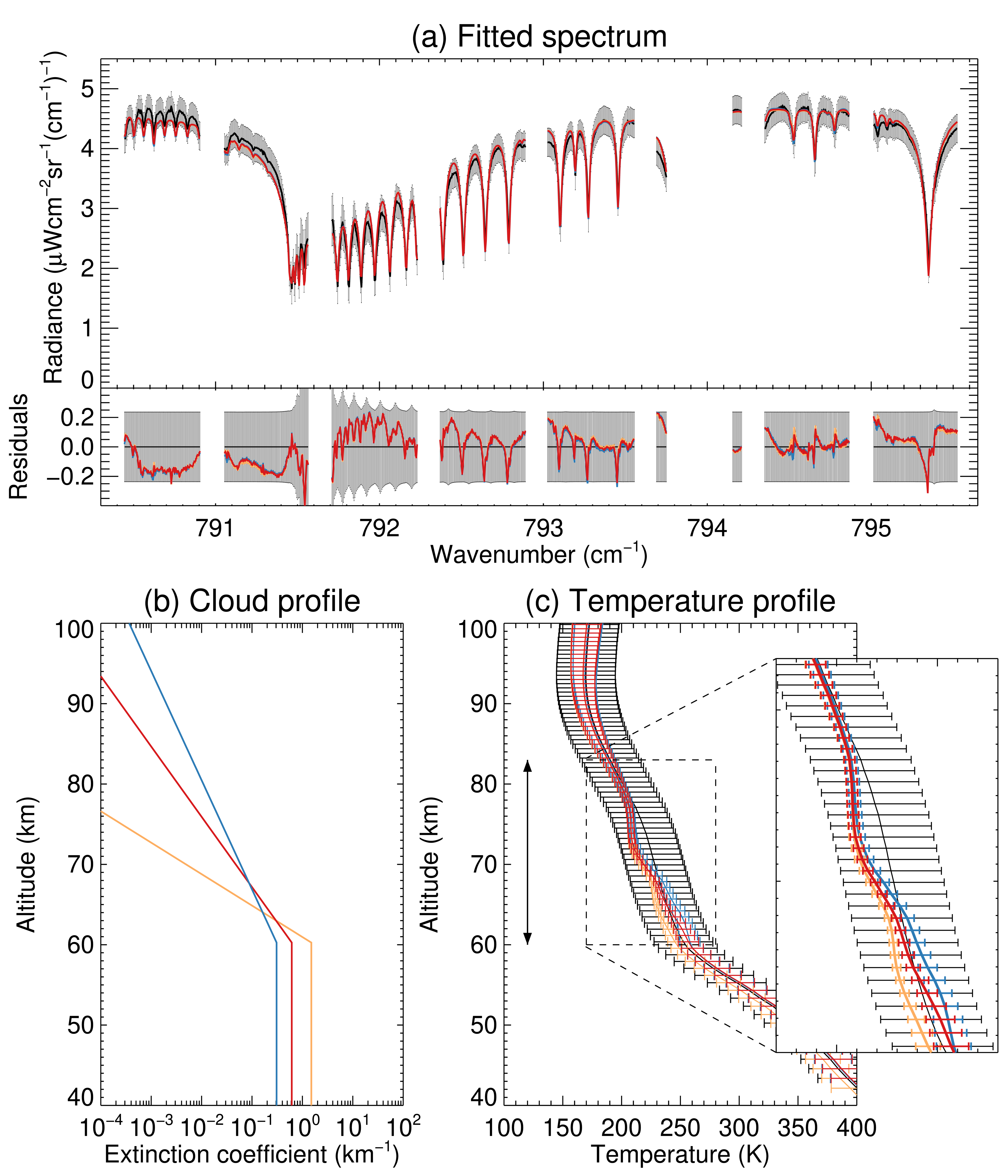}
\caption{An example NEMESIS retrieval of TEXES Venus data, using data obtained on February 12, 2019. The retrieval was run three times using three different cloud model scale heights: 1.7 km (orange), 3.8 km (red) and 5.9 km (blue). (a) The reduced TEXES data (black) along with the best-fit modeled spectra (red, blue and orange). The residuals show the difference between the modeled spectra and the data. The error bars on the data combine the measurement error and the modeling error (see Section~\ref{sec:spectral_modeling}). The small gaps in the spectrum are due to the gaps between the spectral orders and the larger gap at 794 cm\textsuperscript{-1} is due to the removal of a region containing a strong telluric water line. (b) The best-fit vertical cloud structure obtained for each retrieval. (c) The a priori vertical temperature profile (black) along with the retrieved best-fit vertical temperature profiles (red, blue, orange). The vertical arrows highlight the region of peak sensitivity.}
\label{fig:fitted_spectrum}
\end{figure}

Figure~\ref{fig:fitted_spectrum} shows an example NEMESIS retrieval for one spatial point on Venus. In order to understand the effect of our assumed cloud model, which used a fixed scale height of 3.8 km~\cite[the best fit value at low latitudes,][]{lee12}, we also ran retrievals using a fixed scale heights of 1.7 km~\cite[the best fit value in the polar regions,][]{lee12} and 5.9 km. All three retrievals are presented in Figure~\ref{fig:fitted_spectrum}; the H = 3.8 km model is shown in red, the H = 1.7 km model is shown in orange and the H = 5.9 km model is shown in blue.

Figure~\ref{fig:fitted_spectrum}(a) shows the TEXES spectrum in black, along with the best-fit model spectra in red, blue and orange. All three cloud models are able to produce good fits to the data. Figure~\ref{fig:fitted_spectrum}(b) shows the retrieved vertical cloud profile in each case. Even though each model has a different scale height, they all have a similar cloud top level, defined as the altitude where the optical depth in the 790-796 cm\textsuperscript{-1} interval reaches one; the retrieved cloud top levels are 62.1 km, 63.7 km and 64.0 km for H = 1.7 km, H = 3.8 km and H = 5.9 km respectively. The retrieved quantity we are primarily interested in, the vertical temperature profile, is shown in Figure~\ref{fig:fitted_spectrum}(c). The a priori temperature profile is shown in black, while the three retrieved temperature profiles are shown in red, blue and orange. As noted above, the TEXES spectra are primarily sensitive to the 60--83 km altitude range (highlighted by the vertical arrows), and this is therefore the region where the retrieved temperature profiles have the smallest error bars (2--6 K). Figure~\ref{fig:fitted_spectrum}(c) shows that the three retrieved temperature profiles are the same above $\sim$68 km but differ slightly at greater depths. In the following discussion, we therefore bear in mind that although the data is sensitive to down to 60 km, the 60--68 km range is somewhat susceptible to cloud-related degeneracies.  

Using our initial assumption of a cloud with a scale height of 3.8 km, we repeated the retrieval process for each spatial point in the latitude-local time grid, in order to build up a three-dimensional temperature map of Venus. These maps were then converted into temperature offset maps by subtracting the mean zonal temperature at each latitude and altitude level. These three-dimensional temperature offset maps are discussed in Section~\ref{sec:results}.

\section{Results}
\label{sec:results}

\begin{figure}[p]
\centering
\includegraphics[width=12cm]{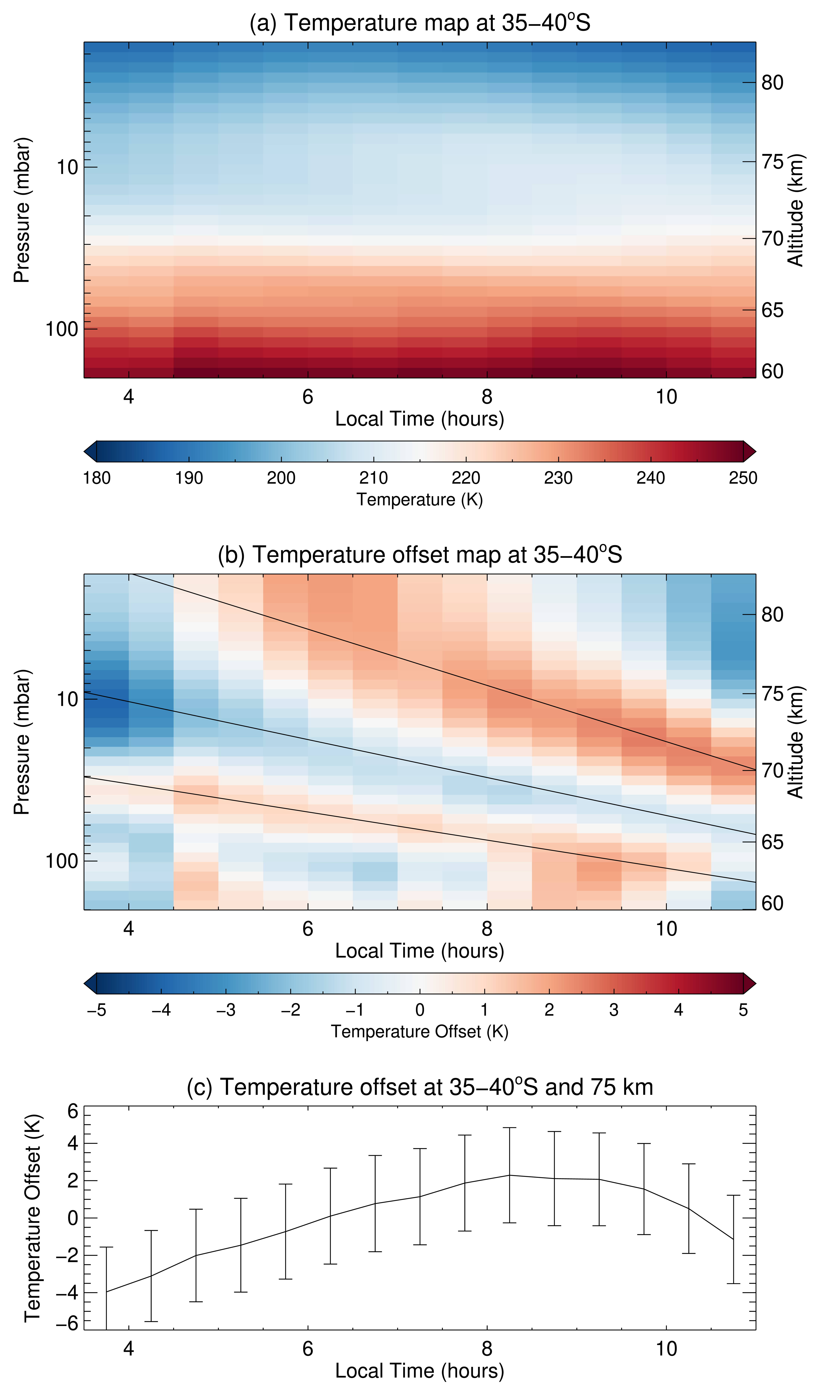}
\caption{(a) A vertical slice through the three-dimensional temperature map of Venus at a latitude of 35--40$^{\circ}$S, obtained on February 12, 2019 (b) The same data as (a) but presented instead as a temperature offset map. For each altitude level, the temperature offset is defined as the temperature relative to the mean zonal temperature. The black lines highlight the gradients of the banded structure. (c) A single altitude level from (b), showing the error bars on the retrieved temperature offsets.}
\label{fig:localtime_altitude}
\end{figure}

Figure~\ref{fig:localtime_altitude}(a) shows an example vertical `slice' through the three-dimensional retrieved temperature map of Venus, taken at a latitude of 35--40$^{\circ}$S. Note that because Venus rotates from east to west, increasing local time corresponds to further west longitudes. The overall temperature structure is similar to the a priori vertical temperature profile shown in Figure~\ref{fig:fitted_spectrum}(c) and any superimposed thermal wave pattern is difficult to see. Figure~\ref{fig:localtime_altitude}(b) presents the same retrieved temperatures as Figure~\ref{fig:localtime_altitude}(a), but in the form of a temperature offset map. As described in Section~\ref{sec:spectral_modeling}, the temperature offset is defined as the temperature relative to the zonally-averaged temperature at that latitude and altitude level, i.e., the mean temperature offset for each row in Figure~\ref{fig:localtime_altitude}(b) is zero. Figure~\ref{fig:localtime_altitude}(c) shows a single altitude level from Figure~\ref{fig:localtime_altitude}(b).

\begin{figure}
\centering
\includegraphics[width=12cm]{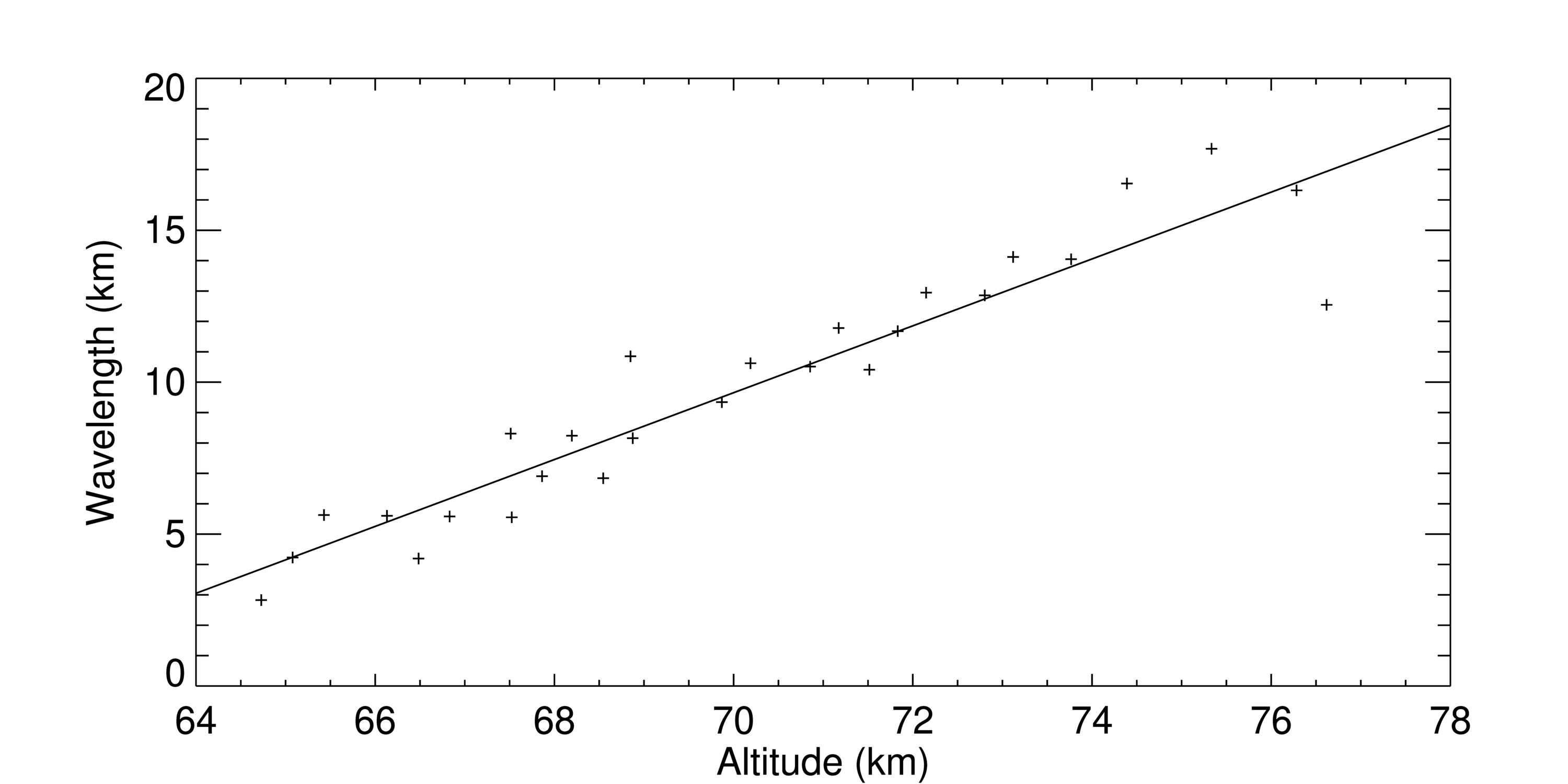}
\caption{Vertical wavelength as a function of altitude for the wave structure shown in Figure~\ref{fig:localtime_altitude}.}
\label{fig:scatter}
\end{figure}

Figure~\ref{fig:localtime_altitude}(b) shows a clear vertical wave structure above $\sim$65 km: a banded pattern that tilts morning-ward (eastward) with altitude. The black lines highlight the three main wavefronts that can be seen; these three lines have gradients of 2.0, 1.4 and 1.0 km per hour, which equate to 8, 11 and 15 degrees per km. The wavefronts do not appear to be perfectly parallel in local time - log(pressure) space; instead, the vertical and horizontal wavelengths appear to vary with altitude. In the horizontal direction, the gap between an adjacent maximum and minimum is 3--5 hours, equivalent to zonal wavenumbers of $\sim$2--4, with lower wavenumbers at higher altitudes. The variation in the vertical wavelength is demonstrated in Figure~\ref{fig:scatter}, which shows the local wavelength (twice the distance between an adjacent maximum and minimum) plotted against altitude (the halfway point between the adjacent maximum and minimum). Within the altitude region we are sensitive to, the vertical wavelength ranges between 5 and 15 km. The peak to peak amplitude of the wave pattern increases with altitude, from $\sim$3 K at 65 km to $\sim$6 K at 75 km. The actual amplitudes may be slightly higher, particularly at high altitude, because the local time range does not necessarily include both a local maximum and a local minimum (e.g. Figure~\ref{fig:localtime_altitude}(c) does not include a local minimum). 

\begin{figure}
\centering
\includegraphics[width=10.5cm]{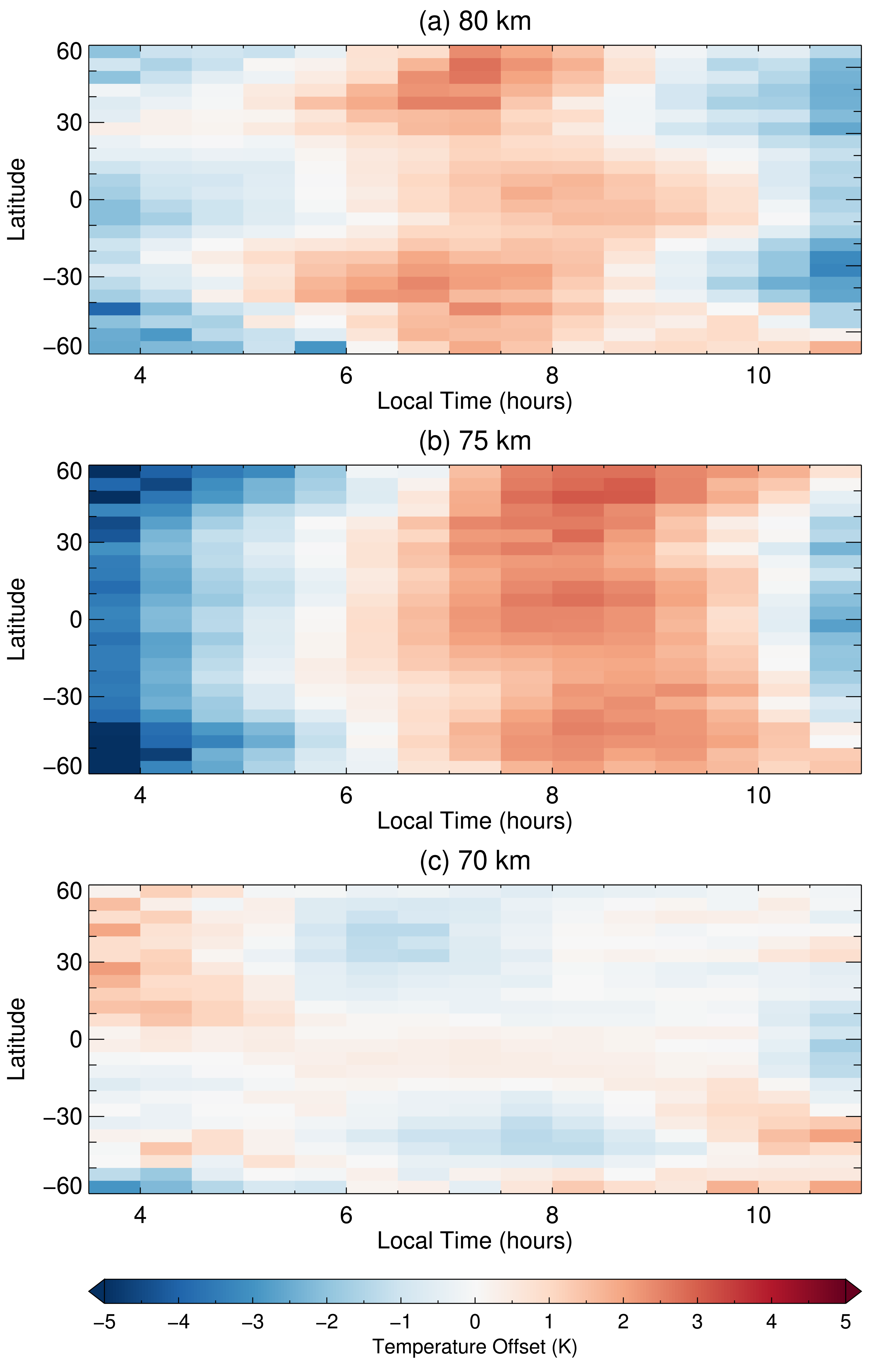}
\caption{Temperature offsets as a function of local time and latitude for three different altitude levels, observed on February 12, 2019. The temperature offset is defined as the temperature relative to the mean zonal temperature.}
\label{fig:localtime_latitude}
\end{figure}

Figure~\ref{fig:localtime_altitude} focuses on a single latitude bin but this pattern persists across a range of latitudes. This can be seen in Figure~\ref{fig:localtime_latitude}, which shows horizontal `slices' through the three-dimensional temperature offset map of Venus at three different altitude levels. The vertical wave pattern is strongest in the mid-latitudes, at 20--55$^{\circ}$. At these latitudes, there is a clear phase difference between the three plots, which matches up with the vertical slice at 35--40$^{\circ}$S shown in Figure~\ref{fig:localtime_altitude}. At 80 km, the temperatures offsets in the mid-latitudes peak at around 7 AM. At 75 km, they peak slightly later, at around 8:30 AM, and by 70 km the peak is at around 10:30 AM, with another peak becoming visible in the northern mid-latitudes at 3:30 AM. In contrast, any phase shift with latitude at the equator is less apparent. Across all three latitudes, the equatorial temperatures peak at around 8 AM; at 80 km this is slightly later than the mid-latitudes, while at 75 km this is slightly earlier. 

\section{Discussion and Conclusions}
\label{sec:discussion}

The wave pattern described in Section~\ref{sec:results} has the following characteristics:

\begin{enumerate}
    \item The wave fronts tilt towards the morningside (eastward) with altitude.
    \item The tilt of the wavefronts is 8--15 degrees per km.
    \item The wave pattern is visible in the mid-latitudes of the planet, at $\sim$20--55$^{\circ}$.
    \item The wave has a zonal wavenumber of 2--4, with lower wavenumbers at higher altitudes.
    \item The vertical wavelength increases with altitude, with wavelengths of 5--15 km in the altitude region we are probing.
    \item The peak to peak amplitude increases with altitude, from $\sim$3 K at 65 km to $\sim$6 K at 75 km.
\end{enumerate}

These ground-based observations are consistent with thermal tide results that have been previously obtained from spacecraft instruments. \cite{schofield83b} used observations from Pioneer Venus OIR to infer a vertical temperature profile between 65 and 100 km. Their results show several similarities with the results obtained from IRTF/TEXES. Below 100 km, they found that the dominant zonal wavenumber was 2 and the wavenumber-2 amplitude increased from 2 K at the cloud tops to 4--5 K at 80 K. This is consistent with our measured amplitudes which range from $\sim$3 K at 65 km to $\sim$6 K at 75 km. They also found that the solar-fixed wavenumber-2 structure tilted eastward with increasing altitude, at a rate of 6 degrees per km, which is slightly steeper than the steepest wavefront shown in Figure~\ref{fig:localtime_altitude} (8 degrees per km). 




More recently, \cite{kouyama19} and \cite{akiba21} used the LIR instrument on the Akatsuki mission to image the global structure of Venus' thermal tides. \cite{kouyama19} found that the semidiurnal (wavenumber of 2) component dominates at latitudes below 40$^{\circ}$, while the diurnal (wavenumber of 1) component dominates at high latitudes. Because the TEXES observations are restricted to the Earth-facing hemisphere and therefore have limited local time coverage, it is difficult to accurately measure the zonal wavenumber; however the gaps between adjacent maxima and minima suggest a dominant wavenumber of $\geq$2 in the mid-latitudes ($\sim$20--55$^{\circ}$). At an altitude of approximately 70$\pm$10 km, the LIR observations show a local peak at $\sim$8--9 AM, which is similar to the results shown in Figures~\ref{fig:localtime_altitude} and \ref{fig:localtime_latitude}. Unlike Pioneer Venus OIR, Akatsuki LIR consists of a single filter, which limits the vertical information. However, by comparing observations obtained at multiple emission angles, \cite{akiba21} were able to reconstruct vertical temperature profiles at altitudes of $\sim$66--71 km and produce two-dimensional slices similar to Figure~\ref{fig:localtime_altitude}(b). They found that the average tilt of the wave pattern across the 66-71 km range is 9.4 degrees per km, but that this decreases to 6.7 degrees per km when only considering the top half of this range. This is consistent with our TEXES observations, which shows tilts ranging from 15 degrees per km at $\sim$66 km to 8 degrees per km at $\sim$75 km.

This eastward-tilted wave structure that has been observed by Pioneer Venus OIR, Akatsuki LIR, and now IRTF/TEXES is consistent with upward motion of an eastward-propagating wave. This is what is expected from Venus' thermal tides; thermal tides are thought to be solar-synchronous, so move eastward along with the sub-solar point and they propagate upwards and downwards from the forcing region in the upper cloud deck~\citep{sanchez-lavega17}. As our observations are sensitive to the region immediately above the cloud tops, upward motion is consistent with expectations. Similar structures have also been reproduced in numerical models of Venus' atmosphere. \cite{takagi18} used a Venus General Circulation Model to investigate the three-dimensional structure of thermal tides and found that wave fronts tilt in opposite directions above and below the cloud layer, indicating upward propagation above 70 km and downward propagation below 60 km. Our observations suggest that the eastward tilt, and therefore upwards propagation, persist down to $\sim$65 km.

The TEXES observations show a vertical wavelength that increases from 5 km at an altitude of 65 km to 15 km at an altitude of 75 km. These values are slightly smaller than those obtained by~\cite{akiba21} using Akatsuki LIR data, who measured a vertical wavelength of $\sim$20 km for both the diurnal and semidiurnal components of the thermal tide. They are broadly consistent with \cite{baker87} who used a numerical model of Venus' atmosphere to investigate the semidiurnal tide and found that the vertical wavelength at 60--75 km could vary from 13--20 km depending on the specific model parameters used. The apparent variation in the vertical wavelength with altitude could be due to the presence of multiple wave components with different wavelengths, or it could be due to the fact that the wavelength depends strongly on the zonal wind and the static stability of the atmosphere, both of which vary significantly with altitude~\citep{takagi18}.

The eastward-titled wave structure with wavenumber $\geq$2 that is observed in the TEXES data is strongest at $\sim$20--55$^{\circ}$; at the equator, there is no clear phase shift with altitude. This is somewhat surprising as numerical models of the thermal tides have suggested that the semidiurnal component of the thermal tides should be strongest at latitudes below 30$^{\circ}$~\citep{takagi18}. \cite{schofield83b} and \cite{akiba21} did observe the tilted wave structure at low latitudes, but in both cases, they averaged data taken over a relatively large latitudinal zone of 0--30$^{\circ}$N. It is possible that this effect, which appears to be confined to within 15$^{\circ}$ of the equator, can only be seen in higher spatial resolution data. Future studies will explore whether the change in wave pattern at the equator is a consistently observed phenomenon. 

This paper presents the first three-dimensional global temperature maps of Venus obtained from ground-based observations. Ground-based observations provide a complementary view to spacecraft instruments and have several distinct advantages. The high spectral resolution that can be achieved with IRTF/TEXES allows the temperature structure to be vertically resolved, while the fact that we are able to observe an entire hemisphere of the planet gives us the ability to obtain range of spatial and local time coverage within a short time period. Ground-based observations also provide significant flexibility, allowing us to continue to make similar observations several times per year over a multi-year period. This allows complete local time coverage to be built up and temporal trends to be observed. Future studies will take advantage of observation dates where Venus' angular diameter is larger in order to obtain better spatial resolution and will combine multiple IRTF/TEXES datasets obtained on different dates in order to study the short- and long-term variability in the three-dimensional thermal tide structure. 

\section*{Acknowledgements}

The authors were Visiting Astronomers at the Infrared Telescope Facility, which is operated by the University of Hawaii under contract 80HQTR19D0030 with the National Aeronautics and Space Administration. This work was funded by NASA Solar System Observations Grant 80NSSC20K0669. 

\printcredits

\bibliographystyle{cas-model2-names}

\bibliography{main.bib}

\end{document}